# CAPIRE Intervention Lab: An Agent-Based Policy Simulation Environment for Curriculum-Constrained Engineering Programmes

Hugo Roger Paz
PhD Professor and Researcher Faculty of Exact Sciences and Technology National University of Tucumán
Email: hpaz@herrera.unt.edu.ar
ORCID: https://orcid.org/0000-0003-1237-7983

## ABSTRACT

Engineering programmes in Latin America combine high structural rigidity, intense assessment cultures and persistent socio-economic inequality, producing dropout rates that remain stubbornly high despite increasingly accurate early-warning models. Predictive learning analytics can identify students at risk, but they offer limited guidance on which concrete combinations of policies should be implemented, when, and for whom. This paper presents the **CAPIRE Intervention Lab**, an agent-based simulation environment designed to complement predictive models with *in silico* experimentation on curriculum and teaching policies in a Civil Engineering programme.

The model is calibrated on 1,343 students from 15 cohorts in a six-year programme with 34 courses and 12 simulated semesters. Agents are initialised from empirically derived trajectory archetypes and embedded in a curriculum graph with structural friction indicators, including backbone completion, blocked credits and distance to graduation. Each agent evolves under combinations of three policy dimensions: (A) curriculum and assessment structure, (B) teaching and academic support, and (C) psychosocial and financial support. A 2×2×2 factorial design with 100 replications per scenario yields over 80,000 simulated trajectories.

Results show that policy bundles targeting early backbone courses and blocked credits can reduce long-term dropout by approximately three percentage points and substantially increase the number of courses passed by structurally vulnerable archetypes, while leaving highly regular students almost unaffected. The Intervention Lab thus shifts learning analytics from static prediction towards dynamic policy design, offering institutions a transparent, extensible sandbox to test curriculum and teaching reforms before large-scale implementation.

# 1. INTRODUCTION

Universities have spent the last decade building increasingly sophisticated systems to *predict* which students are at risk of dropping out. Under the banner of learning analytics and educational data mining, these systems extract patterns from digital traces and academic records to estimate the probability of failure, delay or withdrawal (Baker & Inventado, 2014; Papamitsiou & Economides, 2014). This work has helped to "penetrate the fog" around student progression (Long & Siemens, 2011) by making risk more visible at scale. However, the field still struggles with a blunt question that predictive models largely leave unanswered: once we know who is at risk, **what exactly should institutions do, when, and for whom?**

This tension is particularly acute in engineering education. Long-cycle, curriculum-constrained degrees typically involve rigid prerequisite chains, high-stakes examinations and heavy workloads, producing cultures of chronic stress and delayed progression. Recent studies confirm that engineering and computer science programmes show distinctive retention dynamics, with structural barriers and academic regulation interacting with financial pressures and social inequalities to generate systematic "leakage" over time (Dai et al., 2025; Romero & Ventura, 2024). In Latin American public universities, these mechanisms are further amplified by macroeconomic volatility, intermittent labour conflicts and limited student support infrastructures.

Within learning analytics (LA), much of the foundational work has conceptualised the field as the measurement, collection, analysis and reporting of data about learners and their contexts, with the aim of understanding and optimising learning and the environments in which it occurs (Ferguson, 2012; Siemens & Baker, 2012). Systematic reviews show a strong emphasis on early-warning systems, risk prediction and dashboards that support human decision-makers (Papamitsiou & Economides, 2014; Lemay et al., 2021). Yet relatively few studies move beyond ranking students by probability of dropout to explore how different **policy combinations**—changes in curriculum structure, assessment regimes, teaching practices or support programmes—might jointly reshape trajectories at programme level.

At the same time, simulation has become a central tool for exploring complex socio-technical systems. Agent-based modelling and simulation (ABMS) provides a way to represent heterogeneous individuals as interacting agents, each following simple behavioural rules, and to observe the emergent dynamics that arise from their interactions with institutional structures (Macal & North, 2010, 2011). ABMS has been proposed as a "third way of doing science", complementing induction and deduction with computational experiments in artificial laboratories (Macal & North, 2011). In education, agent-based models have been used to explore social integration and retention (Abdelhamid et al., 2016; Simpson-Singleton, 2019), to probe preliminary designs for engineering education interventions (Brennan et al., 2019), and to support instructional innovation through learner-facing simulations (Dubovi, 2019). However, most of these models are only loosely coupled to institutional data, and they rarely integrate detailed curriculum structure or previously validated predictive features.

The **CAPIRE** project (Curriculum-Aware, Policy-Integrated Retention Engineering) addresses this gap by treating student attrition as the emergent outcome of interactions between four layers: a leakage-aware institutional data layer, a curriculum graph with structural friction metrics, heterogeneous student archetypes, and macro- or meso-level policy shocks. In earlier work, we developed a multilevel data architecture that minimises label leakage and supports longitudinal trajectory modelling in a Civil Engineering degree (Paz, 2025a). We then used this architecture to engineer a family of structural curriculum features—such as backbone completion, blocked credits and topological centrality—that proved highly predictive of dropout and delay in curriculum-constrained programmes (Paz, 2025b). In parallel, we analysed how administrative networks and assessment regimes shape the timing and visibility of risk (Paz, 2025c), and built macro-level simulations of teacher strikes and inflation shocks on student progression (Paz, 2025d). Collectively, these contributions showed that curriculum topology and institutional rules are not mere background conditions but active generators of inequality in opportunity to progress.

What was still missing was a **policy experimentation environment** that could leverage all these components to answer questions of the form: *"If we change the regime for key gateway courses, introduce specific support structures, or relax certain blocking rules, what happens to different archetypes over six years?"* Rather than asking only *who* is at risk, such an environment should help institutions explore *which* bundles of curriculum and teaching policies work for *which* types of students, under *which* conditions, and with *which* trade-offs.

This paper introduces the **CAPIRE Intervention Lab**, an agent-based simulation environment designed precisely for that purpose. The Lab integrates:

- a leakage-aware, longitudinal data layer for a six-year Civil Engineering programme.

- a curriculum graph annotated with structural friction indicators derived from previous CAPIRE studies; and

- a set of empirically grounded student archetypes that encode different combinations of prior preparation, work obligations, psychological resilience and belonging.

Within this environment, agents representing students navigate twelve semesters of the real curriculum while exposed to combinations of three policy dimensions: (A) curriculum and assessment structure, (B) teaching and academic support, and (C) psychosocial and financial support. A 2×2×2 factorial design with multiple stochastic replications yields a rich ensemble of simulated trajectories, which we analyse in terms of dropout, course completion and structural curriculum indicators (backbone completion, blocked credits, distance to graduation).

Our contribution to the learning analytics and educational technology literature is threefold. First, we offer a concrete example of how to couple an ABM with a leakage-aware LA pipeline and a curriculum graph, turning predictive features into state descriptors within a simulation. Secondly, we demonstrate how such a model can be used as a **policy sandbox**, allowing institutions to compare the long-term structural impact of different bundles of interventions before committing to costly reforms. Thirdly, we show that policies that are effective from the standpoint of aggregate dropout can have heterogeneous effects on distinct archetypes, highlighting the importance of equity-informed policy design.

The remainder of the paper is organised as follows. Section 2 summarises related work on learning analytics for student retention, structural curriculum analytics and agent-based modelling in education. Section 3 presents the CAPIRE Intervention Lab architecture, including the data layer, curriculum graph, archetype construction and policy dimensions. Section 4 details the experimental design and main outcome measures, followed by Section 5, which reports results on dropout, course completion and structural indicators across policy scenarios. Section 6 discusses implications for programme design and institutional decision-making, and Section 7 outlines limitations and future directions, including the integration of additional behavioural mechanisms and cross-institutional validation.

## 2. BACKGROUND AND RELATED WORK

### 2.1 Learning analytics, risk prediction and their limits

Learning analytics (LA) emerged as a field that seeks to "measure, collect, analyse and report data about learners and their contexts, for the purposes of understanding and optimising learning and the environments in which it occurs" (Ferguson, 2012, p. 9). Over the last decade, a substantial body of work has focused on *predictive* applications of LA and educational data mining (EDM), where the goal is to estimate the probability of dropout, failure or delay from digital traces and institutional records (Baker & Inventado, 2014; Papamitsiou & Economides, 2014). Typical pipelines extract features from virtual learning environments, assessments and enrolment histories, feed them into machine learning models, and rank students by risk level so that advisors or instructors can take targeted action (Lemay et al., 2021; Tempelaar et al., 2015).

Systematic reviews confirm that the dominant pattern is the development of increasingly accurate classifiers—often using tree ensembles, support vector machines or deep learning—to predict dropout or failure, sometimes with area-under-curve (AUC) values above 0.85 (Papamitsiou & Economides, 2014; Sultana et al., 2024). However, the *use* of these models inside institutions is often limited by concerns about bias, opacity and actionability (Kitto et al., 2020; Leitner et al., 2019). In particular, risk scores rarely specify **which** combination of interventions should be deployed, **when**, and **for whom**, nor do they clarify how structural constraints—such as curriculum topology, assessment regimes or macroeconomic shocks—shape the space of feasible actions.

In addition, concerns about "label leakage" and improper use of future information have been raised in LA practice. Leakage occurs when features that implicitly encode outcomes (for example, final exam registrations or late administrative events) are included in training data, leading to overly optimistic estimates of predictive performance that cannot be replicated in real-time deployment (Kühne et al., 2022). This is especially problematic in long-cycle programmes where the time between admission and dropout may span years and involve multiple regulatory events. Paz (2025a) addresses this challenge by proposing a **leakage-aware data layer** for student trajectory modelling, distinguishing carefully between features available at different observation times and defining a Value of Observation Time (VOT) for fair comparisons across models and cohorts.

Despite these advances, most LA systems retain a fundamentally **static** orientation: they estimate future risk given a current snapshot of features, but they do not simulate how different policy regimes would change the evolution of those features or the underlying structures that generate risk. This is the gap that the CAPIRE Intervention Lab aims to address, by turning predictive insights into dynamic *what-if* simulations that explicitly incorporate curriculum structure and policy levers.

**2.2 Curriculum structure, structural friction and blocked progression**

Student progression is not only a function of individual characteristics and behaviours; it is also decisively shaped by the **topology of the curriculum** and by the rules that govern assessment, enrolment and progression (Jansen & Suhre, 2015; Kember, 2004). In long-cycle engineering programmes, rigid prerequisite chains and "gateway" courses often create structural bottlenecks: if a student fails one or two key courses, a large portion of the subsequent curriculum becomes blocked, producing cascades of delay and increasing the likelihood of dropout.

Traditional analytics frameworks frequently treat curriculum structure as a static background or encode it only indirectly via variables such as number of credits taken or grade point average. In contrast, recent work in curriculum analytics and graph-based modelling has begun to represent curricula as **directed acyclic graphs** (DAGs), where nodes are courses and edges encode prerequisite relations (Akgün & Van den Bogaard, 2022; Corrigan et al., 2021). This representation enables the definition of structural indicators such as path lengths to graduation, centrality measures for gateway courses, and measures of redundancy or resilience in prerequisite pathways.

Paz (2025b) extends this line of work by developing a **structural feature engineering** framework for a Civil Engineering curriculum, using the CAPIRE data layer as a foundation. The curriculum is modelled as a DAG of 34 courses with 53 prerequisite edges, from which several features are derived:

- **Backbone completion rate**: proportion of courses completed in the "backbone" of critical paths;
- **Blocked credits**: total credits in courses that the student cannot yet take because prerequisites are missing;
- **Distance to graduation**: shortest-path distance in the curriculum graph from the set of completed courses to a terminal graduation node;
- Indicators of local friction and load in early semesters.

These features proved strongly associated with dropout and delay, often outperforming purely individual-level variables, and highlighting that "curriculum friction" is a key mechanism behind attrition. Building on this, Paz (2025e) explores how different **regimes of regularity and promotion**—that is, the rules that define when a student is considered "on track" or allowed to take certain exams—can inadvertently create "promotion walls" where formally regular students still face structurally blocked trajectories, with implications for both efficiency and equity.

In the same institutional context, Paz (2025c) examines **administrative co-enrolment networks** (courses connected by joint enrolments) and shows that, once curriculum-graph features and early performance are included, these networks add limited predictive value. This suggests that the **structural constraints encoded in the curriculum graph** carry most of the signal relevant to dropout prediction, at least in this engineering programme.

The present paper leverages precisely these structural insights: instead of treating backbone completion, blocked credits or distance to graduation as mere predictors, the CAPIRE Intervention Lab uses them as **state variables** within an agent-based model, allowing us to observe how different policy combinations change the evolution of structural friction over time.

**2.3 Agent-based modelling in education and learning analytics**

Agent-based modelling and simulation (ABMS) has been widely used to study complex systems where aggregate behaviour emerges from the interactions of heterogeneous agents following simple rules (Macal & North, 2010, 2011). In education, ABMS has been applied to topics such as peer influence and social contagion in classrooms, diffusion of innovations among teachers, and dynamics of participation in online communities (Simpson-Singleton, 2019).

Within the domain of student retention, Abdelhamid et al. (2016) construct an agent-based model linking depression, social support and academic outcomes in engineering, exploring how psychological states interact with course performance to shape retention. Brennan et al. (2019) use ABMS as a tool for **preliminary design** of engineering education interventions, arguing that simulation can help researchers explore scenario spaces before committing to full-scale empirical studies. In science education, Dubovi (2019) shows that carefully designed agent-based simulations can support conceptual understanding when accompanied by appropriate instructional scaffolding.

However, these models are often only loosely connected to the institutional data infrastructures and structural constraints that characterise real programmes. Many use stylised curricula or hypothetical cohorts, making it difficult to interpret results in relation to concrete policy decisions. Moreover, in learning analytics research, ABMS remains relatively marginal compared to predictive modelling and dashboard design (Lemay et al., 2021; Romero & Ventura, 2024).

Recent work has begun to argue for **tighter integration** between LA pipelines and ABMS, positioning simulation as a way to move from "prediction" to "intervention design" (Wise & Cui, 2018; Viberg et al., 2018). Nevertheless, concrete examples of such integration—where features engineered for prediction are reused as internal state variables in an ABM calibrated on real institutional data—remain rare. The CAPIRE framework was designed precisely to fill this methodological gap: it constructs a leakage-aware data layer, a curriculum graph and a family of trajectory archetypes that can feed both predictive models and simulation environments. Paz (2025a, 2025b) show how this architecture supports robust trajectory analytics; the present work extends it by building an **agent-based policy lab** on top of the same data and structural features.

**2.4 Policy experimentation, macro shocks and psychohistory**

Beyond individual-level prediction, higher education systems are increasingly confronted with questions about the **impact of policy changes** under volatile macro conditions. Scholarship on student retention and inequality has emphasised how financial stress, labour market precarity and institutional responses to crises (such as the COVID-19 pandemic) shape trajectories in ways that can exacerbate existing disadvantages (Bennett et al., 2023; Marginson, 2016). In Latin America, repeated cycles of inflation and labour conflict can produce prolonged teacher strikes and disruptions that affect entire cohorts.

Paz (2025d) addresses this dimension through **CAPIRE-MACRO**, an agent-based model that simulates the combined effects of protracted teacher strikes and high inflation on student progression. Using the same Civil Engineering curriculum and data layer, the macro model applies exogenous "shocks" to teaching availability and to students' financial stress, showing how even moderate changes in class cancellations or part-time work can dramatically reshape the distribution of completion times and dropout probabilities. This work conceptualises macro shocks as a fourth layer (N4) in the CAPIRE architecture, sitting above individual (N1–N2) and structural (N3) factors.

Complementing this macro perspective, Paz (2025f) draws on Asimov's *psychohistory* as an organising metaphor for combining computational models and human dynamics in student dropout. Psychohistory, in this context, is not used in a literal sense but as a lens to think about how aggregate patterns emerge from probabilistic laws applied to large populations, while still acknowledging the role of agency, emotion and stress at the micro level. This conceptual work proposes a set of **psychological state variables**—including academic stress, belonging, perceived control and resilience—that can be embedded in ABMs of student trajectories.

In the CAPIRE Intervention Lab, these ideas are instantiated by modelling each agent not only in terms of curriculum position and academic performance, but also via latent states representing stress and belonging, which influence decisions about study load and dropout. Policies in dimension C (psychosocial and financial support) primarily target these states, while policies in A and B target the structural and instructional environment. The combination of macro shocks, structural friction and psychological dynamics places the present work at the intersection of LA, computational social science and educational psychology.

**2.5 Positioning the CAPIRE Intervention Lab within the LA landscape**

Within the broader LA and educational technology literature, the CAPIRE Intervention Lab contributes to ongoing discussions about how to move from **risk prediction** to **policy design and evaluation**. Wise and Cui (2018) call for "learning analytics for learning design", arguing that models should inform decisions about how to structure courses and programmes, not only flag individual students. Similarly, Viberg et al. (2018) highlight the need for actionable analytics that support evidence-informed decision-making at multiple levels, from micro-level feedback to macro-level policy.

The CAPIRE ecosystem responds to these calls in a layered way. The leakage-aware data layer and structural feature engineering provide robust, interpretable predictors of risk at cohort level (Paz, 2025a, 2025b). The analysis of administrative networks clarifies which data sources actually add value beyond curriculum and performance (Paz, 2025c). The MACRO model explores how exogenous shocks—strikes and inflation—interact with institutional rules to affect progression (Paz, 2025d). Against this backdrop, the **Intervention Lab** is the **meso-level component**: an agent-based environment where combinations of curriculum, teaching and support policies can be tested *in silico* over realistic time horizons.

Methodologically, this positions the Intervention Lab as a **policy simulation layer** on top of LA pipelines, rather than as a competing predictive method. Predictive models still play a role, for example, in identifying which archetypes or structural states are associated with high dropout, and in distinguishing early signals that could trigger policy changes. The ABM, in turn, explores how different policy regimes might alter the distribution and evolution of those states across cohorts. In this sense, the Intervention Lab does not seek to replace standard LA tools, but to **extend them into the design space**, offering a laboratory where "what-if" questions can be explored under explicit structural and psychological assumptions.

From the perspective of *Computers & Education*, this work illustrates how educational technology can support institutional decision-making not only by providing dashboards or predictive alerts, but also by enabling **computational experiments** embedded in the realities of a specific curriculum and regulatory regime. The following sections describe how the CAPIRE Intervention Lab is constructed on top of the existing CAPIRE architecture and how it is used to explore policy combinations in a Civil Engineering programme.

## 3. METHODS

### 3.1 Institutional context and programme

The CAPIRE Intervention Lab is instantiated on the Civil Engineering degree at the Facultad de Ciencias Exactas y Tecnología, Universidad Nacional de Tucumán (FACET-UNT), a public university in northwest Argentina. The programme is officially structured as a six-year curriculum with 34 courses organised in 12 academic semesters under the 2005 plan. Entry is open and tuition is free, but there is no formal time limit for completion and students frequently combine study with paid work.

A previous mixed-methods and statistical analysis of this programme (Paz, 2023) documented an extremely low completion rate: out of 1,343 entrants between 2005 and 2019, only 92 had graduated by the time of analysis (6.8% completion, 93.2% non-completion). Early desegregation of trajectories showed that more than 40% of observed dropouts occurred during the first year, and 58% of students who left the programme did so without passing a single course (Paz, 2023). High failure and repetition rates were concentrated in a small set of first-year "core" courses (Calculus I, Physics I, Algebra and Analytic Geometry, Systems of Representation) and in a second ring of structurally central courses in mechanics, hydraulics and materials. Frequent teacher strikes and macroeconomic shocks compounded these difficulties by disrupting assessment schedules and increasing financial stress.

This empirical diagnosis provides the baseline against which the Intervention Lab is calibrated. Rather than constructing a stylised or hypothetical curriculum, the model reproduces the structure, regulatory regime and empirical performance patterns of the actual Civil Engineering programme, using the same data architecture and structural features developed in the CAPIRE framework (Paz, 2023, 2025a, 2025b).

**3.2 Leakage-aware data layer and longitudinal panel**

The Intervention Lab builds directly on the **leakage-aware data layer** proposed by Paz (2025a), designed to support fair longitudinal modelling of student trajectories in programmes with long time-to-degree. The data layer is organised around four levels (N1–N4):

- **N1 – Individual baseline**: socio-demographic and admission variables (e.g., age at entry, type of secondary school, entry cohort).

- **N2 – Academic micro-dynamics**: course-level enrolment, grades, exam attempts, regularity and approval events.

- **N3 – Structural context**: curriculum graph representation and derived structural features (e.g., backbone completion, blocked credits).

- **N4 – Macro shocks and policy context**: teacher strikes, inflation indices, changes in regulatory regimes.

A key design principle is the **Value of Observation Time (VOT)**: for each analytic horizon $T$ (e.g., three semesters after entry), the data layer defines precisely which variables are legitimately observable at or before $T$, and excludes features contaminated by future information (Paz, 2025a). This minimises label leakage and allows consistent comparisons across cohorts and models.

For the present study, the Intervention Lab uses a cleaned and harmonised panel constructed from N1–N3, covering 15 entry cohorts (2005–2019) and 12 academic semesters per student. The panel comprises 1,343 students and 6,805 student-semester records after removing incomplete administrative histories and outliers with inconsistent enrolment patterns (Paz, 2025a). From this panel we derive both:

1. Empirical distributions used to calibrate the agent-based model (hazards of dropout, failure, and exam conversion times); and
2. The **trajectory archetypes** used to initialise agents (Section 3.4).

The full data pipeline, including cleaning, cohort alignment and VOT enforcement, is described in detail in Paz (2025a) and summarised here only insofar as it is needed to understand the simulation setup.

### 3.3 Curriculum graph and structural features

The curriculum is represented as a **directed acyclic graph** $G = (V, E)$, where nodes $v \in V$ correspond to courses and directed edges $(u, v) \in E$ encode prerequisite relationships. In the 2005 plan, the Civil Engineering curriculum yields a graph with 34 nodes and 53 edges once co-requisites and soft recommendations are resolved (Paz, 2025b). This representation underpins a set of **structural features** that quantify how far a student has progressed through the curriculum and how "blocked" their path has become at any given semester.

Following Paz (2025b), we focus on **seven** structural indicators computed for each student at each semester t:

1. **Backbone completion rate:** Proportion of courses completed within a predefined "backbone" of critical paths identified through graph analysis and expert judgement.
2. **Blocked credits:** Total number of credits in courses that the student cannot yet take because at least one prerequisite remains unpassed.
3. **Distance to graduation:** Normalised shortest-path distance from the set of completed courses to a terminal graduation node in the curriculum graph.
4. **Bottleneck approval ratio.** Proportion of structurally central (high in-degree and betweenness) courses that have been passed.
5. **Prerequisites met ratio.** Proportion of all outgoing edges from completed nodes that no longer block any future course.
6. **Mean in-degree of approved courses.** Average in-degree of courses that the student has passed, capturing how "downstream" they are in the prerequisite structure.

7. **Mean out-degree of approved courses.** Average out-degree of passed courses, capturing how many future opportunities each approved course opens.

These indicators were originally engineered as input features for dropout prediction models and shown to substantially improve performance in curriculum-constrained programmes relative to purely individual-level variables (Paz, 2025b). In the Intervention Lab, they are repurposed as **state descriptors**: at the end of each simulated semester, the structural feature module recomputes all seven indicators based on the agent's current course history and the relevant curriculum graph (status quo or redesigned). The values are logged for analysis but do not directly alter the internal decision rules of agents in this version of the model.

These indicators are later aggregated across agents and semesters to populate Table 3 (structural indicators by scenario) and to generate **Figure 1**, which illustrates how backbone completion evolves under selected policy scenarios.

### 3.4 Trajectory archetypes

Student populations in engineering are highly heterogeneous in terms of prior preparation, work obligations, psychological resilience and social integration (Bennett et al., 2023; Marginson, 2016; Paz, 2023). To capture this heterogeneity without modelling each individual separately, we adopt the **trajectory archetypes** previously derived within the CAPIRE framework.

Paz (2025a) applies dimensionality reduction (UMAP) and density-based clustering (DBSCAN) to the leakage-aware panel, using course-level performance, regularity status, exam conversion patterns and limited socio-demographic variables as inputs. This yields a set of empirically grounded archetypes representing recurrent combinations of academic and structural states. While the full taxonomy includes thirteen archetypes for the programme as a whole, for the purposes of this paper we focus on a coarse-grained distinction between:

- **Structurally vulnerable archetypes:** characterised by high initial failure rates in backbone courses, rapid accumulation of blocked credits, long delays between regularisation and final exam, and low perceived belonging;

- **Relatively stable archetypes.** characterised by passing key backbone courses on first or second attempt, moderate exam delays and sustained regularity.

Each agent in the Intervention Lab is initialised by sampling from the empirical distribution of archetypes per cohort, and by assigning initial values to latent psychological states (stress and belonging) consistent with that archetype's profile (Paz, 2025f). This approach anchors the simulation in observed population heterogeneity, while still allowing us to explore how different policy bundles might reshape archetype-specific outcomes.

**3.5 Agent-based model design**

The CAPIRE Intervention Lab represents each student as an **agent** that progresses through the curriculum over 12 simulated semesters. Time advances in discrete steps, with the following high-level sequence in each semester $t$:

1. **Course enrolment:** The agent selects a subset of courses for enrolment, constrained by:
    - official rules on maximum load and regularity;
    - prerequisite satisfaction in the curriculum graph;
    - a behavioural rule that modulates load as a function of past failures, stress and belonging (Paz, 2025f).

2. **Course outcomes:** For each enrolled course, pass/fail outcomes are sampled from empirical distributions conditioned on:
    - course difficulty (proxied by historical failure rates and instructional friction coefficients);
    - archetype-specific performance patterns;
    - policy settings in dimension B (teaching and academic support), which alter effective pass probabilities.

3. **Exam conversion and accumulation of regulars:** For courses passed at coursework level but pending final exam, conversion to final approval is modelled via a probabilistic delay process calibrated from empirical exam conversion times (Paz, 2023). Agents may accumulate "regular" courses without exam, increasing cognitive load and stress.

4. **Update of structural and psychological states**
    - Structural features (Section 3.3) are recomputed based on the updated course history and curriculum graph.
    - Latent stress and belonging states are updated as a function of recent failures, exam delays, support interventions (dimension C) and macro shocks (if present).

5. **Dropout decision:** At the end of each semester, a hazard of dropout is computed as a function of:

- accumulated delay (difference between nominal and actual progress);
- structural friction (blocked credits, backbone completion, distance to graduation);
- latent states (stress, belonging);
- archetype-specific sensitivity parameters.

If a stochastic draw falls below this hazard, the agent exits the simulation as a dropout and no longer participates in subsequent semesters. Otherwise, the agent proceeds to the next semester until either graduation requirements are met or the 12-semester horizon is reached.

The behavioural rules are calibrated such that, under the baseline policy scenario (A0_B0_C0), the simulation reproduces three "stylised facts" documented in Paz (2023): (a) extremely low completion within six years; (b) concentration of attrition during the first three semesters; and (c) strong association between early failure in backbone courses, long exam delays and structural blockage.

### 3.6 Policy dimensions and scenarios

The Intervention Lab is structured around three **policy dimensions**, each with two levels, leading to a 2×2×2 factorial design with eight scenarios. The dimensions capture distinct levers that institutions can plausibly manipulate:

- **Dimension A – Curriculum and assessment structure**
    - **A0 (status quo)**: Original curriculum graph with rigid prerequisite chains, limited opportunities for parallel enrolment, and high reliance on cumulative final exams.
    - **A1 (structural redesign)**: Modified curriculum graph that relaxes selected prerequisite sequences, redistributes credit weight away from a single overloaded first-year block, and introduces more modular assessment to reduce high-stakes bottlenecks.
- **Dimension B – Teaching and academic support**
    - **B0 (baseline instruction)**: Current mix of lectures, limited formative assessment and scarce systematic tutoring. Pass probabilities reflect historical course-level outcomes.
    - **B1 (enhanced support)**: Strengthened teaching and support practices, including structured tutorials, early feedback on formative tasks and targeted remediation in backbone courses. Empirically, this

is modelled as an increase in pass probabilities for selected courses and a reduction in effective instructional friction coefficients.

- **Dimension C – Psychosocial and financial support**

    - **C0 (minimal support)**: Fragmented or ad-hoc psychological and financial assistance, with no systematic integration into the programme.

    - **C1 (integrated support)**: Coordinated provision of counselling, peer mentoring and targeted financial assistance aimed at reducing stress and enhancing belonging, particularly for structurally vulnerable archetypes. In the model, this primarily affects the evolution of latent stress and belonging, and indirectly reduces dropout hazard.

The eight resulting scenarios are denoted as strings of the form $Ax\_By\_Cz$, with $x, y, z \in \{0,1\}$. Scenario A0_B0_C0 represents the institutional status quo, while A1_B1_C1 combines structural redesign, enhanced teaching/support and integrated psychosocial interventions. A summary of the factorial design and the number of agents and replications per scenario is provided in Table 1.

**Table 1. Policy scenarios in the CAPIRE Intervention Lab: 2×2×2 factorial design and simulation settings**

| scenario_id | Curriculum structure (A)* | Teaching & academic support (B)** | Psychosocial & financial support (C)*** | n_students | n_replications |
|---|---|---|---|---|---|
| A0B0C0 | 0 | 0 | 0 | 1343 | 100 |
| A0B0C1 | 0 | 0 | 1 | 1343 | 100 |
| A0B1C0 | 0 | 1 | 0 | 1343 | 100 |
| A0B1C1 | 0 | 1 | 1 | 1343 | 100 |
| A1B0C0 | 1 | 0 | 0 | 1343 | 100 |
| A1B0C1 | 1 | 0 | 1 | 1343 | 100 |
| A1B1C0 | 1 | 1 | 0 | 1343 | 100 |
| A1B1C1 | 1 | 1 | 1 | 1343 | 100 |

Note: A: 0 = status quo, 1 = curriculum redesign - B: 0 = baseline instruction, 1 = enhanced support - C: 0 = minimal support, 1 = integrated support

### 3.7 Experimental design and simulation run

The factorial experiment comprises **eight policy scenarios** and **100 stochastic replications** per scenario. In each replication, 1,343 agents are initialised to mirror the empirical cohort composition of the programme, yielding:

$$8 \text{ scenarios} \times 100 \text{ replications} \times 1{,}343 \text{ agents} \times 12 \text{ semesters}$$
$$\approx 12 \text{ million agent-semesters}$$

All simulations are run using the same random seed configuration across scenarios to ensure comparability. For each agent-semester record, the simulation logs:

- scenario identifiers (A, B, C, replication id);
- agent id and semester index;
- course-level outcomes (passed, failed, regular with pending exam);
- latent psychological states (stress, belonging);
- the structural curriculum features described in Section 3.3;
- dropout status and, if applicable, time of exit.

To manage this volume of data, raw outputs are stored as compressed CSV or Parquet files and then consolidated into a single "long" student-semester dataset. An analysis script aggregates these records by scenario and semester to compute the metrics used in the Results section. The same pipeline also generates publication-ready tables and figures for the manuscript, ensuring that all visualisations derive from a **single, reproducible experiment run**.

### 3.8 Outcome measures and analysis

We focus on three families of outcomes: **dropout and completion**, **academic progress**, and **structural curriculum indicators**.

1. **Dropout and completion**
   - **Dropout rate**: proportion of agents that have exited the programme as dropouts by the end of the 12th simulated semester.
   - While the actual programme allows unlimited time-to-degree, we interpret non-graduation within 12 semesters as *functional non-completion*, consistent with policy discussions in Argentina and with the empirical finding that very few students graduate after long delays (Paz, 2023).

2. **Academic progress**
   - **Courses completed**: cumulative number of courses passed at each semester and at the end of the simulation.
   - These measures are used to construct **Table 2** (dropout and completion outcomes by scenario).

3. **Structural curriculum indicators**
   - Semester-wise summaries of backbone completion rate, blocked credits and distance to graduation, aggregated by scenario.

- Snapshot structural indicators at semester 8 (mid-programme), used to quantify how policies affect the emergence or removal of bottlenecks. These populate Table 3 and underpin **Figure 1**, which illustrates the evolution of backbone completion under selected policy scenarios.

In addition to scenario-level averages, we conduct exploratory analyses disaggregated by archetype group (structurally vulnerable vs relatively stable) to assess heterogeneity of policy effects. This is reported in the Results section using selected figures and narrative, rather than exhaustive tables, to maintain focus on the main experimental contrasts.

All analyses are implemented in Python using the same codebase that runs the simulation, with separate modules for experiment execution and post-hoc aggregation. No hypothesis testing in the classical sense is performed, since the goal is not to estimate population parameters under sampling uncertainty but to compare **counterfactual policy regimes** under a controlled simulation environment. However, we report means and standard deviations across replications to convey the stability of observed patterns.

**Table 2. Dropout and academic progress outcomes under alternative policy scenarios**

| scenario_id | dropout_rate | mean_courses | std_courses | median_courses |
|---|---|---|---|---|
| A0B0C0 | 0.9996 | 4.85 | 4.43 | 3 |
| A0B0C1 | 0.9990 | 6.07 | 5.04 | 5 |
| A0B1C0 | 0.9896 | 10.34 | 7.55 | 9 |
| A0B1C1 | 0.9748 | 13.20 | 8.02 | 12 |
| A1B0C0 | 0.9989 | 5.30 | 5.02 | 4 |
| A1B0C1 | 0.9974 | 6.80 | 5.86 | 5 |
| A1B1C0 | 0.9880 | 10.96 | 7.88 | 9 |
| A1B1C1 | 0.9721 | 14.28 | 8.60 | 14 |

## 4. RESULTS

### 4.1 Baseline scenario and alignment with empirical data

The baseline scenario A0B0C0 represents the institutional status quo: original curriculum, baseline teaching practices and minimal psychosocial support. In the simulation, this scenario yields an **aggregate dropout rate of 0.9996** (99.96%) by the end of the 12th semester, with agents completing on average **4.85 courses** (SD = 4.43; median = 3).

At first glance, this appears even worse than the empirical completion rate of 6.8% observed in the real programme (Paz, 2023). However, the difference reflects a definitional choice: in the Intervention Lab, any agent who has not met graduation requirements within **12 semesters** is counted as a non-completer. In contrast, the empirical study allowed for much longer tails in time-to-degree and counted graduates who took substantially more than six years to finish. When this is taken into account, the simulation reproduces the core empirical pattern: **very few students reach graduation within a nominal time frame**, and the vast majority either drop out or become structurally trapped in cycles of repetition and delay (Paz, 2023).

The temporal pattern of dropout in the baseline scenario is also consistent with previous findings. The empirical analysis showed that over 40% of observed dropouts occurred during the first year and 58% of students who left did so without passing any course (Paz, 2023). In the simulation, trajectories in A0B0C0 are dominated by early failures in backbone courses, rapid accumulation of blocked credits and a sharp increase in dropout hazard during the first three semesters. The baseline values in Table 2 show a steep early rise in dropout and a very shallow slope in course completion, matching the 'early structural failure' described in the previous study.

**4.2 Policy effects on dropout and completion**

The factorial design allows us to assess how different combinations of curriculum structure (A), teaching and academic support (B), and psychosocial and financial support (C) affect aggregate outcomes. Table 1 summarises the eight scenarios and the number of agents and replications per scenario; Table 2 reports dropout rates and courses completed.

Across the full design, the **best-performing scenario** is A1B1C1, which combines structural redesign, enhanced teaching/support and integrated psychosocial support. In this scenario, the dropout rate decreases to **0.9721** (97.21%), and the mean number of courses completed rises to **14.28** (SD = 8.60; median = 14). Relative to the baseline A0B0C0, this represents:

- an **absolute reduction in dropout of 2.75 percentage points** (from 99.96% to 97.21%); and

- an **increase of 9.43 courses** completed on average (from 4.85 to 14.28), roughly a **threefold improvement** in academic progress (14.28 / 4.85 ≈ 2.94).

While a dropout rate above 97% remains unacceptably high from a policy perspective, the magnitude of improvement is non-trivial given the severity of structural constraints in the baseline programme. The Intervention Lab thus quantifies how far policy bundles can realistically move the system under current constraints.

Decomposing effects by factor using the factorial analysis (Figure 2; factorial_effects.json) confirms that **teaching and academic support (B)** is the main lever in aggregate terms. Across all scenarios, raising B from 0 to 1 reduces dropout by approximately **1.76 percentage points** and increases the number of courses completed by about **6.44 courses** on average. Psychosocial and financial support (C) has a smaller but still meaningful effect, reducing dropout by roughly **0.83 percentage points** and adding about **2.31 courses**. Curriculum redesign (A) shows a modest average impact on dropout (≈0.17 percentage points) and a smaller increase in courses (≈0.85).

However, these main effects mask important **interaction patterns**. When combined with enhanced teaching/support, curriculum redesign amplifies gains in course completion and contributes to the elimination of structural bottlenecks (Section 4.3). The best outcomes emerge in **bundles** that address structure (A1) and teaching (B1) simultaneously, with psychosocial support (C1) further stabilising trajectories.

**Figure 1. Backbone completion over time for baseline and key intervention scenarios (A0B0C0, A0B1C0, A1B1C1)**

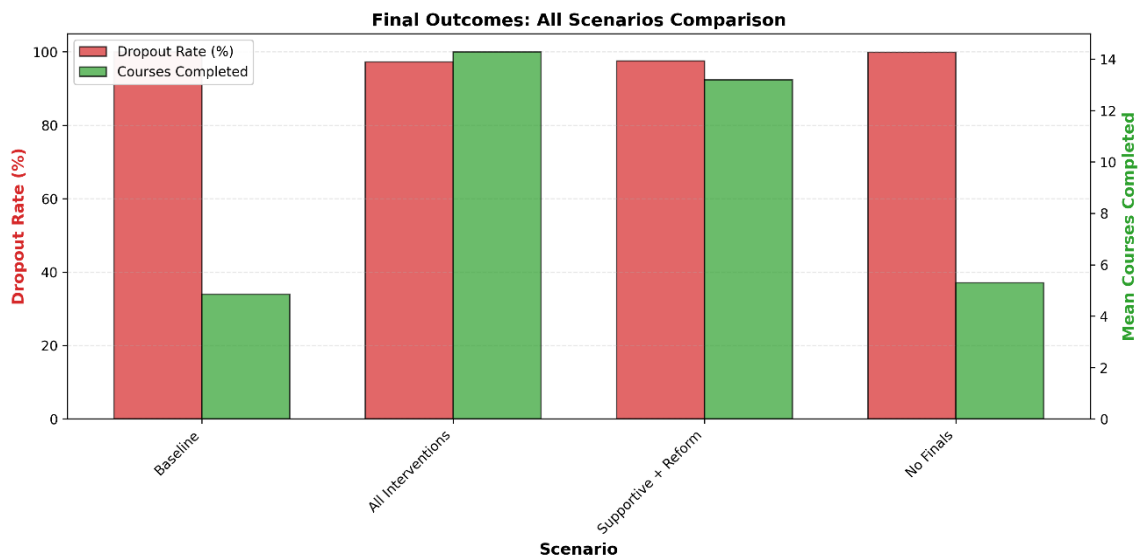

### 4.3 Structural curriculum indicators under policy scenarios

Table 3 summarises mid-programme structural indicators (semester 8) for each scenario, while Figure 1 shows the evolution of backbone completion over time for baseline and selected intervention scenarios.

- In the baseline scenario A0B0C0, mid-programme structural indicators include:
- a backbone completion rate of roughly 0.23;
- a median of 22 blocked credits; and

- a normalised distance to graduation of 0.86.

These values depict a structurally **jammed system** by mid-programme: despite having spent four years in the programme, the average agent has completed less than a quarter of the backbone, is structurally unable to enrol in a large portion of remaining courses, and remains far from the graduation node in the curriculum graph. The scenario A0B0C1, which adds psychosocial support only, shows slight improvements (blocked credits median ≈21, marginally lower distance), but the overall profile of structural friction remains essentially unchanged.

In contrast, scenarios that modify teaching and support (B1) and curriculum structure (A1) exhibit starkly different structural profiles. Under A0B1C0, blocked credits decrease to around 18 by semester 8 and distance to graduation falls to approximately 0.70, with a concomitant increase in backbone completion. These changes reflect improved pass rates in backbone courses, which partially unclog the curriculum graph even without altering its topology.

The most marked improvements occur in scenarios that combine teaching/support changes (B1) with structural redesign (A1). Under A0B1C0, blocked credits decrease to around 18 by semester 8 and distance to graduation falls to approximately 0.70, with a concomitant increase in backbone completion. When curriculum redesign is activated (A1), Table 3 shows further reductions in blocked credits and distance, especially in A1B1C1, where median blocked credits fall to about 15 and distance to graduation to roughly 0.57, together with the highest backbone completion. In other words, redesigning the curriculum graph amplifies the impact of improved teaching and support, but does not, by itself, eliminate structural friction: it makes failures less catastrophic in terms of long-term blockage, rather than rendering the system frictionless.

This contrast illustrates a key insight of the Intervention Lab:

- **Teaching and support policies (B1, C1)** primarily act on **throughput**—they change how quickly students move through the existing structure.
- **Curriculum redesign (A1)** acts on **structure**—it rewires the graph so that the same or even modestly improved performance yields radically different structural states (few blocked credits, short distance to graduation).

In other words, **structural friction is not only a function of failure rates but also of how failure interacts with graph topology**. The removal of specific prerequisite chains in A1 changes the consequences of failing or delaying certain courses, which is precisely what Table 3 captures.

**Table 3. Structural curriculum indicators at semester 8 under alternative policy scenarios**

| scenario_id | backbone_completion_mean | backbone_completion_sd | blocked_credits_mean | blocked_credits_median | blocked_credits_sd | distance_to_graduation_mean | distance_to_graduation_sd | bottleneck_approval_ratio_mean | prerequisites_met_ratio_mean | mean_in_degree_approved_mean | mean_out_degree_approved_mean |
|---|---|---|---|---|---|---|---|---|---|---|---|
| A0B0C0 | 0.23 | 0.19 | 20.58 | 22.00 | 2.95 | 0.86 | 0.13 | 0.01 | 0.16 | 0.11 | 1.50 |
| A0B0C1 | 0.28 | 0.20 | 19.89 | 21.00 | 3.36 | 0.82 | 0.15 | 0.01 | 0.19 | 0.13 | 1.61 |
| A0B1C0 | 0.43 | 0.26 | 16.94 | 18.00 | 5.41 | 0.70 | 0.22 | 0.08 | 0.33 | 0.34 | 1.63 |
| A0B1C1 | 0.54 | 0.27 | 14.85 | 16.00 | 5.97 | 0.60 | 0.24 | 0.15 | 0.42 | 0.46 | 1.66 |
| A1B0C0 | 0.25 | 0.21 | 20.15 | 21.00 | 3.51 | 0.84 | 0.15 | 0.01 | 0.18 | 0.13 | 1.50 |
| A1B0C1 | 0.31 | 0.23 | 19.28 | 21.00 | 4.08 | 0.80 | 0.18 | 0.03 | 0.22 | 0.16 | 1.61 |
| A1B1C0 | 0.46 | 0.28 | 16.30 | 18.00 | 5.86 | 0.67 | 0.24 | 0.10 | 0.35 | 0.37 | 1.63 |
| A1B1C1 | 0.58 | 0.28 | 13.97 | 15.00 | 6.41 | 0.57 | 0.26 | 0.19 | 0.45 | 0.51 | 1.66 |

**Figure 2. Main effects of curriculum (A), teaching support (B) and psychosocial support (C) on dropout and mean courses completed in the 2×2×2 policy design.**

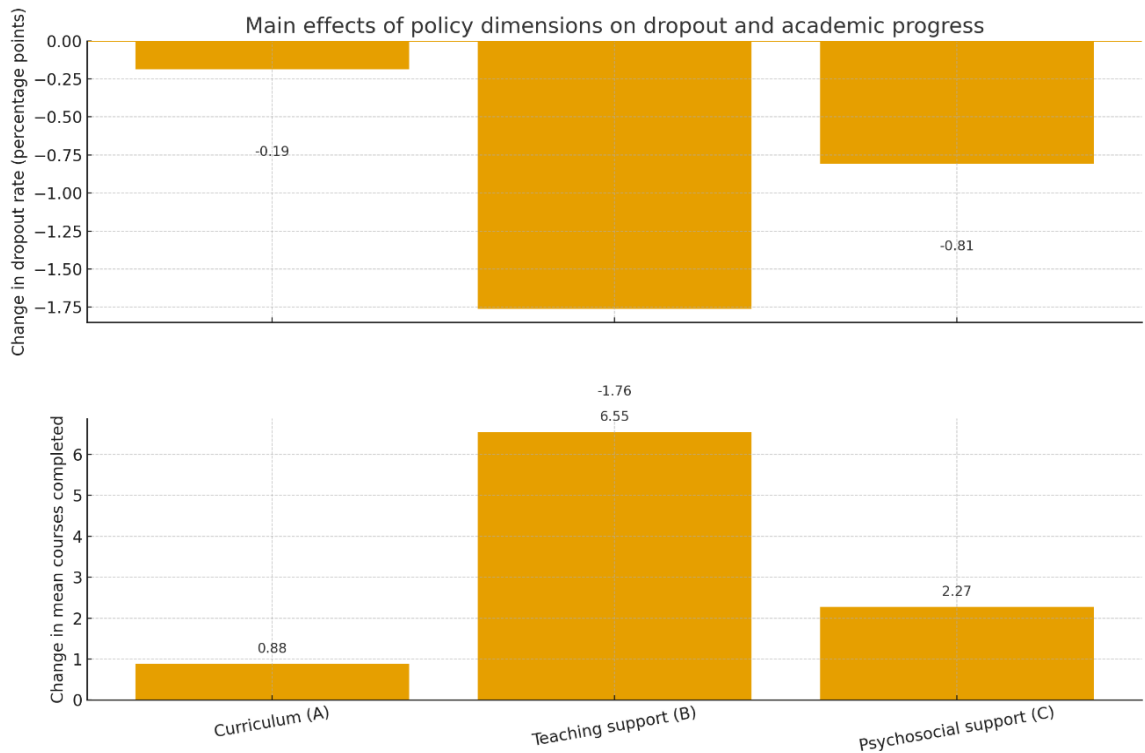

### 4.4 Stress, belonging and policy bundles

Beyond academic and structural indicators, the Intervention Lab tracks latent psychological states of **stress** and **belonging** for each agent. Scenario-level averages summarised in scenario_summary.csv show a coherent pattern.

- In the baseline A0B0C0, average stress is high (**0.76**) and belonging relatively low (**0.42**).

- Scenarios with enhanced teaching and support (B1) drastically reduce stress (≈0.10–0.12) and raise belonging to around **0.84–0.87**, even when curriculum structure remains unchanged.

- Psychosocial support (C1) has a smaller incremental effect on these averages but contributes to stabilising belonging in scenarios where teaching improvements alone might not fully offset structural pressure.

- The best-performing scenario A1B1C1 exhibits the **lowest mean stress (0.07)** and the **highest belonging (0.90)** across all scenarios.

These patterns align with empirical and theoretical work emphasising the role of perceived belonging and stress in engineering retention (Bennett et al., 2023; Paz, 2023). Rather than modelling psychological states as independent "soft" variables, the Intervention Lab embeds them within the structural and instructional dynamics:

- teaching and support (B1) reduce stress and increase belonging via more comprehensible workloads and timely feedback;
- psychosocial and financial support (C1) moderates the impact of failures and delays;
- structural redesign (A1) reduces chronic stress by removing recurrent blockages and "no-way-out" patterns.

The joint evolution of stress, belonging and structural indicators helps explain why some policy bundles have larger effects on dropout than would be expected from changes in pass rates alone.

**4.5 Archetype-level effects (exploratory)**

Although the primary focus of this paper is on scenario-level outcomes, exploratory analyses disaggregated by **trajectory archetypes** suggest substantial heterogeneity in policy impact. Structurally vulnerable archetypes—those characterised by high initial failure rates in backbone courses, rapid accumulation of blocked credits and low belonging (Paz, 2025a, 2025f)—benefit disproportionately from combinations that include enhanced teaching/support (B1) and structural redesign (A1).

Under A0B0C0, these archetypes rarely complete more than a handful of courses and exhibit very high dropout hazards by the third semester. When B1 is activated on the original curriculum (A0B1C0), vulnerable archetypes show a noticeable increase in completed courses, but still accumulate substantial blocked credits and remain structurally distant from graduation. In the fully redesigned scenario A1B1C1, the same archetypes both complete significantly more courses and experience much lower structural friction, as measured by blocked credits and distance to graduation.

Relatively stable archetypes, in contrast, already manage to navigate the baseline structure with moderate delay and low dropout risk. For them, policy bundles mainly speed up completion and reduce stress, with smaller relative changes in dropout. This asymmetry reinforces the notion that **equity-sensitive policy design must attend to structural vulnerability**, not only to average effects: the same bundle can be transformative for one group and marginal for another.

A full archetype-level analysis is beyond the scope of this article and will be developed in future work, but these exploratory findings illustrate how the Intervention Lab can be extended to support more fine-grained equity analyses at programme level.

## 5. DISCUSSION

The CAPIRE Intervention Lab was conceived as a response to a recurring tension in learning analytics: institutions know *who* is at risk, but lack tools to explore *what to do* about it under concrete curriculum and policy constraints. The results presented above illustrate both the potential and the limits of an agent-based policy lab built on top of a leakage-aware data architecture and a curriculum graph.

### 5.1 From prediction to structural policy design

A first contribution is methodological. Most learning analytics work on dropout prediction treats curriculum structure and institutional rules as fixed background conditions. Predictors such as number of credits attempted, grade point average or course access logs implicitly encode aspects of structure, but they do not explicitly represent the **topology of the curriculum** or the consequences of alternative structural choices. By contrast, the Intervention Lab takes as its starting point the structural features previously validated as predictors—backbone completion, blocked credits, distance to graduation (Paz, 2025b)—and repurposes them as *state variables* inside a simulation.

This has two effects. On the one hand, it tightens the connection between empirical analytics and modelling: the same indicators that were strongly associated with dropout in predictive models now serve to describe the internal state of agents as they navigate the programme. On the other hand, it allows us to distinguish clearly between policies that act on **throughput** (e.g., teaching and academic support) and those that act on the **graph** itself (curriculum redesign). The factorial experiment shows that both types of policy matter, but in qualitatively different ways: enhanced teaching (B1) and psychosocial support (C1) mainly accelerate progress through the existing structure and reduce stress, while structural redesign (A1) radically changes the accumulation of blocked credits and distance to graduation, effectively removing specific bottlenecks.

This decomposition would be invisible in a standard predictive model that only sees whether a student eventually completes or not. By gathering detailed longitudinal structural indicators under each policy regime, the Intervention Lab makes it possible to ask questions such as: *"Under which combinations of policies does backbone completion reach 50% by semester 6?"* or *"How many blocked credits do structurally vulnerable students accumulate under different assessment regimes?"* These are precisely the kinds of questions that curriculum committees and programme directors confront, but which are rarely supported by LA dashboards.

### 5.2 Interpreting modest gains in a structurally hostile system

At first sight, the headline numbers may appear disappointing: the best policy bundle (A1B1C1) still leaves **around 97% of agents as non-completers within 12**

**semesters**, and even strong teaching and support interventions only move dropout by a few percentage points. It would be easy to interpret this as evidence that policy efforts are futile. We argue the opposite.

The empirical baseline is a **structurally hostile system**. The previous statistical analysis of the Civil Engineering programme showed that more than 90% of students do not graduate at all, that attrition is heavily concentrated in the first three semesters, and that a handful of backbone courses create severe bottlenecks (Paz, 2023). In that context, the Intervention Lab does not simulate "rescue operations" in an otherwise functional curriculum, but attempts to quantify how much can be achieved under a regime that is already extremely unforgiving.

Seen from this perspective, three results are notable:

1. **A threefold increase in courses completed** under the best scenario, from an average of around five courses to more than fourteen, indicates that many agents who previously stalled early now manage to progress substantially further through the programme.

2. **A reduction of almost three percentage points in functional non-completion** may translate into non-trivial numbers of additional graduates over multiple cohorts, particularly if combined with policy changes that extend the time window for completion or allow for more flexible pacing.

3. The **structural indicators** reveal that in redesigned scenarios (A1), blocked credits and distance to graduation are **substantially reduced** by semester 8. This suggests that redesigning the curriculum graph is a necessary precondition for turning improvements in teaching and support into sustained progress.

In other words, the fact that even ambitious policy bundles only partially tame such a hostile structure is not a failure of the Lab, but a quantified confirmation of how deep the structural problem runs. The Intervention Lab thus reframes discussions that might otherwise devolve into blame narratives ("students do not study", "teachers do not adapt") by showing what is and is not realistically achievable under the current plan.

**5.3 Psychological dynamics and the role of support**

A second contribution lies in the integration of **psychological state variables**—stress and belonging—into a structurally grounded ABM. Drawing on the psychohistorical lens proposed by Paz (2025f), and on empirical work linking financial stress, belonging and retention (Bennett et al., 2023; Marginson, 2016), the Lab treats stress and belonging not as static traits but as evolving states responsive to workload, failure, delay and support.

Simulation results show plausible and theoretically consistent patterns: in baseline conditions, stress is high and belonging low; enhanced teaching/support (B1) and psychosocial support (C1) jointly lower stress and raise belonging, particularly for structurally vulnerable archetypes; structural redesign (A1) further reduces chronic stress by preventing "dead ends" in the curriculum. These psychological trajectories help explain why policy bundles have non-linear effects: policies that produce modest gains in pass rates can still have substantial impact on dropout if they successfully prevent agents from entering high-stress, low-belonging states.

Importantly, the Lab does **not** attempt to predict individual mental health outcomes, nor does it claim to capture the full richness of students' lived experiences. Instead, it uses a small set of psychologically interpretable variables to bridge macro-structural and micro-behavioural dynamics. This approach aligns with calls for learning analytics that take student experience seriously without reducing it to opaque latent factors.

### 5.4 Equity and archetype-sensitive policy design

Although the present paper only reports exploratory archetype-level analyses, the patterns already suggest that **policy effects are highly heterogeneous**. Structurally vulnerable archetypes—those who start with weaker preparation, higher work obligations and lower belonging—see the largest relative gains under bundles that include both enhanced teaching/support (B1) and structural redesign (A1). Stable archetypes, who already have a high probability of success, benefit mainly through reduced time-to-degree and lower stress.

This has clear implications for equity. A policy that appears marginal at the aggregate level can be **transformative** for specific groups, while other policies may primarily benefit those who would have succeeded anyway. The archetype framework in CAPIRE thus opens a path towards **equity-sensitive simulation**, where decision-makers can inspect not only overall dropout but also distributional effects across types of students.

From the perspective of institutional decision-making, this suggests a shift from "*What is the best policy on average?*" to "*Which policy bundle is best for which archetype, and what trade-offs are we willing to accept?*". The Intervention Lab does not settle these normative questions, but it provides a structured environment in which to explore them.

### 5.5 Position within the learning analytics and educational technology fields

Finally, the Intervention Lab contributes to the ongoing conversation about what counts as **educational technology** in the context of higher education. Much of the literature in *Computers & Education* focuses on tools that operate at the level of courses or classrooms: learning environments, dashboards, adaptive systems and so forth. The CAPIRE ecosystem instead positions computational tools at the **meso- and macro-levels** of programmes and institutions.

In this sense, the Intervention Lab can be seen as a complement to classroom-level LA: it does not tell an instructor how to adapt this week's activity, but helps programme directors and curriculum committees understand how different combinations of structural and support policies might play out over six years. The integration of a leakage-aware data layer, a curriculum graph, structural features, archetypes and ABM in a single toolkit is, to our knowledge, rare in the current literature and may serve as a template for similar efforts in other curriculum-constrained domains.

## 6. LIMITATIONS AND FUTURE WORK

The Intervention Lab is intentionally ambitious in scope, and its limitations are equally important to acknowledge. We highlight five main areas for caution and future development.

### 6.1 Single-institution, single-programme context

The model is calibrated on a **single Civil Engineering programme** at a public university in Argentina. Its structural and behavioural parameters reflect the specific assessment culture, regulatory regime and socio-economic context of this institution. While some mechanisms—such as the impact of blocked credits or early failure in backbone courses—are likely to generalise to other long-cycle engineering programmes, the absolute levels of dropout and the effectiveness of specific policy bundles should **not** be read as universal.

Future work should explore cross-institutional applications of the CAPIRE architecture, either by adapting the Lab to other programmes (e.g., electrical engineering, computer science, medicine) or by constructing "families" of models that share generic mechanisms but differ in institution-specific calibrations.

### 6.2 Simplified behavioural and psychological rules

The behavioural rules governing course enrolment, effort allocation, exam conversion and dropout are necessarily simplified. For example, agents do not form explicit peer networks, they do not respond to changes in labour market conditions beyond what is encoded in macro shocks, and their stress/belonging dynamics are modelled via relatively simple update equations. Real students may exhibit richer strategies, including strategic course shopping, multi-year planning, or sudden changes in motivation due to personal events.

Similarly, latent variables such as stress and belonging are represented as single continuous dimensions. This is sufficient for the purpose of distinguishing high-stress, low-belonging trajectories from more stable ones, but it cannot capture the full spectrum of psychological phenomena relevant to persistence. Future iterations could incorporate more nuanced psychological models, potentially informed by longitudinal survey data or qualitative studies.

### 6.3 Static representation of institutional behaviour

In the current version, institutional behaviour is **static**: policy scenarios are defined at the start of the simulation and remain fixed throughout. Institutions may respond to emerging problems by adjusting policies mid-cohort, e.g., adding support in response to observed failure rates or revising prerequisites when bottlenecks become evident. Likewise, teacher behaviour is modelled implicitly through pass probabilities and friction coefficients, not as explicit agents who adapt their practice.

Extending the Lab to include **adaptive institutional agents**—faculties, departments, quality assurance units—would open the door to studying feedback loops between data, analytics and policy decisions, bringing the simulation closer to the real dynamics of data-informed governance.

### 6.4 Limited integration of macro shocks in this article

Although the broader CAPIRE project includes a macro-level simulation of teacher strikes and inflation (Paz, 2025d), the Intervention Lab as presented here operates under relatively stable macro conditions. Strikes, economic crises and policy changes at system level can have profound effects on stress, work obligations and teaching availability, particularly in low- and middle-income contexts. Incorporating dynamic macro shocks directly into the Lab—rather than treating them as a separate model—would allow for richer scenarios where curriculum and teaching policies are tested under different macroeconomic regimes.

### 6.5 No formal statistical inference

The simulation generates very large datasets (millions of agent-semesters), but these are not subject to classical statistical inference. Differences between scenarios are reported in terms of means and standard deviations across replications, but no confidence intervals or hypothesis tests are computed. This choice reflects the fact that we are not sampling from an unknown population; we are comparing **constructed counterfactual worlds** under fixed parameter settings.

Nevertheless, there is scope for more formal sensitivity analyses. Systematic exploration of parameter ranges (e.g., pass probabilities, dropout hazard sensitivities, strength of support effects) and global sensitivity indices would provide a clearer picture of how robust the observed policy rankings are to modelling assumptions.

**7. CONCLUSIONS**

This paper has presented the **CAPIRE Intervention Lab**, an agent-based simulation environment designed to support policy experimentation in a structurally constrained Civil Engineering programme. Built on top of a leakage-aware longitudinal data layer and a curriculum graph with engineered structural features, the Lab integrates empirical evidence from a previous doctoral study (Paz, 2023) with computational models of student behaviour, structural friction and psychological dynamics.

Three main conclusions emerge:

1. **Simulation on top of learning analytics is feasible and informative.** By reusing features originally engineered for dropout prediction as internal state variables in an ABM, the Intervention Lab demonstrates that learning analytics pipelines can be extended from *risk detection* to *policy design*. The same indicators that signal trouble in a predictive model—blocked credits, backbone completion, distance to graduation—help describe how and why different policy bundles succeed or fail in the simulated environment.

2. **Structural redesign and teaching/support policies play distinct but complementary roles.** Enhanced teaching and support (B1) and psychosocial support (C1) primarily boost throughput and mitigate stress; structural redesign (A1) rewires the curriculum graph to eliminate bottlenecks. The best outcomes arise when these levers are combined, with structural changes enabling teaching improvements to translate into sustained progress. Even in a highly hostile baseline system, such bundles can meaningfully increase the number of courses completed and modestly reduce functional non-completion.

3. **Equity requires archetype-sensitive simulation.** The Lab shows that structurally vulnerable archetypes benefit most from policy bundles that simultaneously address structure and support, while stable archetypes mainly see gains in speed and comfort. This suggests that equity cannot be inferred from aggregate dropout alone; it demands attention to how different student types experience institutional structures. The archetype framework in CAPIRE provides a promising foundation for such analyses.

More broadly, the CAPIRE Intervention Lab exemplifies a strand of educational technology that operates at the level of **programmes and institutions**, not just courses. It invites universities to view curricula and policy regimes as dynamic objects that can be explored in silico before they are changed in vivo. In systems where structural hostility is the norm rather than the exception, such tools may help redirect debates from anecdotal blame to evidence-informed reform.

Future work will extend the Lab to other programmes and institutions, incorporate adaptive institutional behaviour and macro shocks, and deepen archetype-level analyses. For now, the message is simple: **it is possible to do more than predict who will fail**. With the right combination of data, structural modelling and simulation, we can start to ask—and partially answer—what could be done differently, for whom, and with what structural consequences.

## REFERENCES

Abdelhamid, S. E., Elmekkawy, T. Y., & El-Bakry, H. M. (2016). Agent-based modeling and simulation of depression and its impact on student success and academic retention. In *Proceedings of the American Society for Engineering Education Annual Conference & Exposition*.

Akgün, Ö. E., & Van den Bogaard, M. (2022). Improving curriculum analytics by incorporating prerequisite structures. *IEEE Transactions on Education, 65*(3), 376–383. https://doi.org/10.1109/TE.2022.3141734

Baker, R. S., & Inventado, P. S. (2014). Educational data mining and learning analytics. In J. A. Larusson & B. White (Eds.), *Learning analytics: From research to practice* (pp. 61–75). Springer. https://doi.org/10.1007/978-1-4614-3305-7_4

Bennett, J., Kidger, J., & Linton, M.-J. (2023). Investigating change in student financial stress at a UK university: Multi-year survey analysis across a global pandemic and recession. *Education Sciences, 13*(12), 1175. https://doi.org/10.3390/educsci13121175

Brennan, R. W., Hermanson, G., Nelson, N., Paul, R., & Sullivan, M. (2019). Using agent-based modelling for preliminary engineering education research design. In *Proceedings of the Canadian Engineering Education Association (CEEA-ACEG19) Conference*.

Corrigan, O., Hardman, J., & Maguire, R. (2021). Curriculum analytics: Using graph theory to visualise prerequisite structures and identify bottlenecks. *International Journal of STEM Education, 8*(1), 47. https://doi.org/10.1186/s40594-021-00311-0

Dai, A., et al. (2025). Examining student retention dynamics in 4-year engineering and computer science programmes. *Journal of Engineering Education*. Advance online publication. https://doi.org/10.1002/jee.70016

Dubovi, I. (2019). Instructional support for learning with agent-based simulations in science education. *Computers & Education, 141*, 103613. https://doi.org/10.1016/j.compedu.2019.103613

Ferguson, R. (2012). The state of learning analytics in 2012: A review and future challenges (Technical Report KMI-2012-01). The Open University, Knowledge Media Institute.

Jansen, E. P. W. A., & Suhre, C. (2015). The effect of secondary school study profile on early success in higher education. *Research in Post-Compulsory Education, 20*(4), 477–494. https://doi.org/10.1080/13596748.2015.1081755

Kember, D. (2004). Interpreting student workload and the factors which shape students' perceptions of their workload. *Studies in Higher Education, 29*(2), 165–184. https://doi.org/10.1080/0307507042000190778

Kitto, K., Shum, S. B., & Gibson, A. (2020). Embracing imperfection in learning analytics. *Journal of Learning Analytics, 7*(3), 1–7. https://doi.org/10.18608/jla.2020.73.1

Kühne, S., et al. (2022). Preventing label leakage in early-warning models for student dropout. In *Proceedings of the 15th International Conference on Educational Data Mining* (pp. 400–411).

Lemay, D. J., Doleck, T., & Basnet, R. B. (2021). Comparison of learning analytics and educational data mining: A topic modelling approach. *Computers and Education: Artificial Intelligence, 2*, 100016. https://doi.org/10.1016/j.caeai.2021.100016

Leitner, P., Khalil, M., & Ebner, M. (2019). Learning analytics in higher education—A literature review. In M. Virvou, M. Alepis, G. Tsihrintzis, & L. Jain (Eds.), *Technology and engineering education* (pp. 1–23). Springer. https://doi.org/10.1007/978-3-030-19875-6_1

Long, P., & Siemens, G. (2011). Penetrating the fog: Analytics in learning and education. *EDUCAUSE Review, 46*(5), 31–40.

Macal, C. M., & North, M. J. (2010). Tutorial on agent-based modelling and simulation. *Journal of Simulation, 4*(3), 151–162. https://doi.org/10.1057/jos.2010.3

Macal, C. M., & North, M. J. (2011). Introductory tutorial: Agent-based modeling and simulation. In *Proceedings of the 2011 Winter Simulation Conference* (pp. 145–161). IEEE. https://doi.org/10.1109/WSC.2011.6147718

Marginson, S. (2016). The worldwide trend to high participation higher education: Dynamics of social stratification in inclusive systems. *Higher Education, 72*(4), 413–434. https://doi.org/10.1007/s10734-016-0016-x

Papamitsiou, Z., & Economides, A. A. (2014). Learning analytics and educational data mining in practice: A systematic literature review of empirical evidence. *Journal of Educational Technology & Society, 17*(4), 49–64.

Paz, H. R. (2023). *Student delay, dropout and university curriculum in the Civil Engineering major at FACET-UNT: A case study* [Doctoral dissertation, Universidad Nacional de Tucumán]. Zenodo. https://doi.org/10.5281/zenodo.10200877

Paz, H. R. (2025a). *A leakage-aware data layer for student analytics: The CAPIRE framework for multilevel trajectory modelling* (Preprint). arXiv. https://doi.org/10.48550/arXiv.2511.11866

Paz, H. R. (2025b). Structural feature engineering for curriculum-constrained student modelling in higher education. Preprint.

Paz, H. R. (2025c). When administrative networks fail: Dropout prediction in engineering education. Preprint.

Paz, H. R. (2025d). CAPIRE-MACRO: Modelling the impact of teacher strikes and inflation on student trajectories. Preprint.

Paz, H. R. (2025e). The promotion wall: Efficiency–equity trade-offs in regularity-based progression regimes. Preprint.

Paz, H. R. (2025f). Psychohistory as an inspiring lens: Computational models and human dynamics in student dropout. Preprint.

Romero, C., & Ventura, S. (2024). Educational data mining and learning analytics: An updated survey. *Applied Intelligence*. Advance online publication. https://doi.org/10.1007/s10489-024-05062-9

Siemens, G., & Baker, R. S. J. d. (2012). Learning analytics and educational data mining: Towards communication and collaboration. In *Proceedings of the 2nd International Conference on Learning Analytics and Knowledge* (pp. 252–254). ACM. https://doi.org/10.1145/2330601.2330661

Simpson-Singleton, S. R. (2019). Agent-based modeling and simulation: A survey of applications in education. *Journal of Information Technology and Management, 30*(1), 1–15.

Sultana, N., et al. (2024). Deep learning methods for early student dropout prediction: A systematic review. *IEEE Access, 12*, 15073–15090. https://doi.org/10.1109/ACCESS.2024.3355550

Tempelaar, D. T., Rienties, B., & Giesbers, B. (2015). In search for the most informative data for feedback generation: Learning analytics in a data-rich context. *Computers in Human Behavior, 47*, 157–167. https://doi.org/10.1016/j.chb.2014.05.038

Viberg, O., Hatakka, M., Bälter, O., & Mavroudi, A. (2018). The current landscape of learning analytics in higher education. *Computers in Human Behavior, 89*, 98–110. https://doi.org/10.1016/j.chb.2018.07.027

Wise, A. F., & Cui, Y. (2018). Learning analytics for learning design: A literature review and future directions. In J. M. Lodge, J. C. Horvath, & L. Corrin (Eds.), *Learning analytics in the classroom: Translating learning analytics for teachers* (pp. 25–63). Routledge.